\documentclass[twocolumn,showpacs,preprintnumbers,amsmath,amssymb]{revtex4-1}

\usepackage{graphicx}
\usepackage{dcolumn}
\usepackage{bm}
\usepackage{hyperref}

\begin{document}


\title{Experimental observation of a transition from amplitude to oscillation death in coupled oscillators} 



\author{Tanmoy Banerjee}
\email{tbanerjee@phys.buruniv.ac.in}
\affiliation{Department of Physics, University of Burdwan, Burdwan 713 104, West Bengal, India.}
\author{Debarati Ghosh}
\affiliation{Department of Physics, University of Burdwan, Burdwan 713 104, West Bengal, India.}


\received{:to be included by reviewer}
\date{\today}

\begin{abstract}
We report the first experimental evidence of an important transition scenario, namely the transition from amplitude death (AD) to oscillation death (OD) state in coupled limit cycle oscillators. We consider two Van der Pol oscillators coupled through mean-field diffusion and show that this system exhibits a transition from AD to OD, which was earlier shown for  Stuart-Landau oscillators under the same coupling scheme [T. Banerjee and D. Ghosh, arXiv:1403.2907, 2014]. We show that the AD-OD transition is governed by the density of mean-field and beyond a critical value this transition is destroyed; further, we show the existence of a  nontrivial AD state that coexists with OD. Next, we implement the system in an electronic circuit and experimentally confirm the  transition from AD to OD state. We further characterize the experimental parameter zone where this transition occurs. The present study may stimulate the search for the practical systems where this important transition scenario can be observed experimentally.
\end{abstract}


\maketitle 

The suppression of oscillation has been attracting the attention of researchers due to its ubiquity in diverse fields like physics, biology, and engineering \cite{kosprep}. In coupled oscillators, there exists two distinct types of oscillation quenching processes, namely  amplitude death (AD) and oscillation death (OD). Although, AD and OD are two structurally different phenomena, their clear distinction has been made only recently in pioneering works reported in Ref.\cite{kosprl,kospre,kosprep} (see Ref.\cite{kosprep} for an extensive review on OD). In AD, all the coupled oscillators populate a common stable steady state that was unstable otherwise and thus gives rise to a stable homogeneous steady state (HSS) \cite{prasad1,*asen,*prasad3}. But, in the case of OD, due to symmetry breaking bifurcation, the oscillators populate different coupling dependent stable steady states resulting in stable inhomogeneous steady states (IHSS). Thus, the stabilization of HSS gives rise to AD, and the stabilization of IHSS gives birth to OD. 

In this context, \citet{kosprl} first proved that, despite of their different origin, AD and OD can simultaneously occur in a diffusively coupled system of oscillators; more significantly, they reported an important transition phenomenon, namely the transition from AD to OD. They established the analogy between this transition and the Turing-type bifurcation \cite{turing} in spatially extended systems. The AD-OD transition in identical Stuart-Landau oscillators is reported in \cite{kospre} for the dynamic \cite{dyn}, and conjugate \cite{karna} coupling schemes. In Ref.~\cite{dana,*dana1}, diverse routes to AD-OD transition have been shown in identical nonlinear oscillators that are coupled diffusively with an additional repulsive coupling link. Recently, the present authors have reported the AD-OD transition in Stuart-Landau oscillators coupled via mean-field diffusion \cite{tanod}. In that paper we have shown that the AD-OD transition is governed by the mean-field density parameter. Also, we have reported a novel  nontrivial AD state that coexists with OD for a certain parameter zone, and which is destroyed by the parameter mismatch.

Although OD and AD have been separately observed previously in a number of different dynamical systems \cite{exptodelec,*exptodchem1,*exptodthermo,asen1,tanchaosad}, to the best of our knowledge, hitherto no experimental confirmation of the predicted AD-OD transition has been reported.

In this paper, we report the {\it first} experimental evidence of AD-OD transition in coupled oscillators. For this we consider two Van der Pol oscillators \cite{vdp} in their stable oscillation mode coupled via mean-field diffusion. The paradigmatic Van der Pol oscillator is widely used for the demonstration and understanding of nonlinear dynamics; further, it has a rich connection with engineering and biological systems \cite{vdpbio1,*vdpbio2,*vdpbio3}. The choice of the mean-field coupling is motivated by the fact that it is one of the important coupling schemes owing to its presence in many natural systems \cite{shino,*st,*de,qstr,*srimali,*ninno,*bec,tanchaosad}. Also, experimental observation of AD-OD transition in this coupling scheme is comparatively easy because, as we showed in Ref.\cite{tanod}, there exists a wide parameter region where OD  is the only existing state, which is in contrast to the other diffusive coupling schemes where, in general, OD is accompanied by limit cycle oscillations \cite{kosepl,*koschaos}. We at first cary out theoretical and numerical analyses to explore the dynamical behaviors of the coupled Van der Pol oscillators and characterize the AD-OD transition. Next, the coupled system is implemented in electronic circuit to experimentally demonstrate the transition. Experimental results show the evidence of AD-OD transition for a wide range of parameter values.

We consider two Van der Pol  (VdP) oscillators interacting through mean-field diffusive coupling; mathematical model of the coupled system is given by
\begin{subequations}
\label{system}
\begin{align}
\label{x1}
\dot{x}_{1,2} &= y_{1,2}+\epsilon\left(Q\overline{X}-x_{1,2}\right),\\
\label{y1}
\dot{y}_{1,2} &= a_{1,2}(1-x{^2}_{1,2})y_{1,2}-x_{1,2}.
\end{align}
\end{subequations}
Here $\overline{X}=\frac{1}{2}\sum_{i=1}^{2}x_i$ is the mean-field of the coupled system. The individual VdP oscillator  shows a near sinusoidal oscillation for small $a_{1,2}$, and relaxation oscillation for large $a_{1,2}$. The coupling strength is given by $\epsilon$;  $Q$ is called the mean-field density parameter that determines the density of mean-field diffusion \cite{qstr,*srimali,tanchaosad}; $ 0\le Q <1$. As the limiting case we take two identical VdP oscillators: $a_{1,2}=a$. From Eq.\eqref{system} we can see that the system has the following fixed points: the origin $(0, 0, 0, 0)$ is the trivial fixed point, and two additional coupling dependent fixed points: (i) (${x_1}^\ast$, ${y_1}^\ast$, $-{x_1}^\ast$, $-{y_1}^\ast$) where ${x_1}^\ast =\frac{{y_1}^\ast}{{\epsilon}}$ and ${y_1}^\ast = \sqrt {\epsilon^2-\frac{\epsilon}{a}}$. (ii) (${x_1}^\dagger$, ${y_1}^\dagger$, ${x_1}^\dagger$, ${y_1}^\dagger$) where ${x_1}^\dagger = \frac{{y_1}^\dagger}{\epsilon(1 - Q)}$ and ${y_1}^\dagger = \sqrt {\epsilon^2(1-Q)^2-\frac{\epsilon(1-Q)}{a}}$.

The eigenvalues of the system at the origin are,
\begin{subequations}
\label{lambda}
\begin{align}
\label{lambda1}
{\lambda}_{1,2} &= \frac{(a-\epsilon)\pm\sqrt{(a+\epsilon)^2-4}}{2},\\
\label{lambda3}
{\lambda}_{3,4} &= \frac{(a-\epsilon(1-Q))\pm\sqrt{(a+\epsilon(1-Q))^2-4}}{2}.
\end{align}
\end{subequations}
From eigenvalue analysis we derive the two pitchfork bifurcation (PB) points PB1 and PB2 emerge at the following coupling strengths:
\begin{subequations}
\label{pb}
\begin{align}
\label{epsapb1}
{\epsilon}_{PB1} &= \frac{1}{a},\\
\label{epsapb2}
{\epsilon}_{PB2} &= \frac{1}{a(1-Q)}.
\end{align}
\end{subequations}
The IHSS, (${x_1}^\ast$, ${y_1}^\ast$, $-{x_1}^\ast$, $-{y_1}^\ast$), emerges at ${\epsilon}_{PB1}$ through a symmetry breaking pitchfork bifurcation. The other nontrivial fixed point (${x_1}^\dagger$, ${y_1}^\dagger$, ${x_1}^\dagger$, ${y_1}^\dagger$) comes into existence at ${\epsilon}_{PB2}$, which gives rise to an unique {\it nontrivial HSS} (to be discussed later in this paper). From \eqref{lambda} we can see that no Hopf bifurcation of trivial fixed point occurs for $a>1$; in that case, only pitchfork bifurcations exist. Thus for $a>1$ no AD, and AD-OD transition are possible. For any $a<1$, equating the real part of ${\lambda}_{1,2}$ and ${\lambda}_{3,4}$ to zero, we get two Hopf bifurcation points at 
\begin{subequations}
\begin{align}
\label{epsahb1}
{\epsilon}_{HB1} &= a,\\
\label{epsahb2}
{\epsilon}_{HB2} &= \frac{a}{1-Q}.
\end{align}
\end{subequations}
\begin{figure}
\includegraphics[width=.49\textwidth]{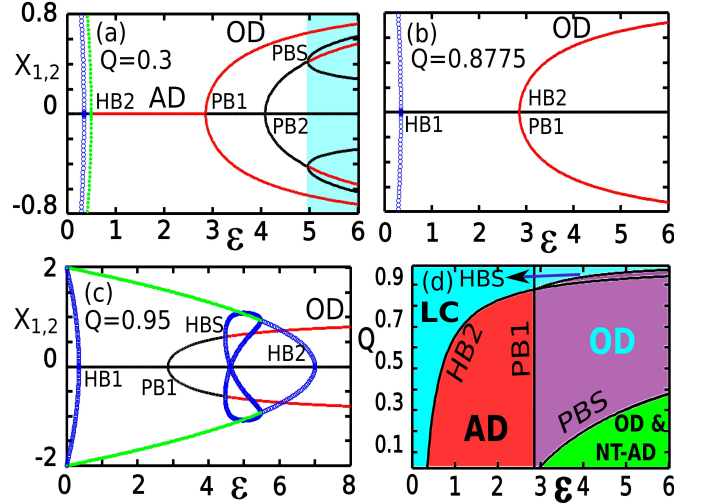}
\caption{\label{fig1}(Color online) Bifurcation diagram (using $\mbox{XPPAUT}$) of two mean-field coupled identical Van der Pol oscillators ($a=0.35$). Grey (red) lines: stable fixed points, black lines: unstable fixed points, solid circle (green): stable limit cycle, open circle (blue): unstable limit cycle. HB1,2 and HBS are Hopf bifurcation points; PB1,2 and PBS are pitchfork bifurcation points. (a) $Q=0.3$ ($<Q^\ast$): AD-OD transition takes place; coexistence of OD ($x_1=-x_2$) and {\it nontrivial} AD (NT-AD) ($x_1=x_2$) is also shown (shaded (cyan) region). (b) $Q=0.8775$ (=$Q^\ast$): AD vanishes, AD-OD transition just destroyed. (c) $Q=0.95$ ($>Q^\ast$): no AD-OD transition, only OD. (d) Phase diagram in $\epsilon-Q$ space. With increasing $Q$, collision of HB2 and PB1 destroys the AD-OD transition scenario.}
\end{figure}
We use $\mbox{XPPAUT}$ package \cite{xpp} to compute the bifurcation branches. Figure~\ref{fig1}~(a) shows the bifurcation diagram of $x_{1,2}$ with $\epsilon$ for $Q=0.3$ and $a=0.35$ [without any loss of generality, throughout this paper, we use $a=0.35$ ($a<1$)].  It is observed that at $\epsilon_{HB2}=0.5$, AD is born through an inverse Hopf bifurcation; whether at $\epsilon_{HB1}=0.35$, an unstable limit cycle is born. This AD (stable HSS) state becomes unstable trough a supercritical pitchfork bifurcation (PB1) to give birth to  OD (stable IHSS) at ${\epsilon}_{PB1}= 1/a=2.857$. Now, with increasing $Q$ value, ${\epsilon}_{HB2}$ moves towards ${\epsilon}_{PB1}$, and at a particular $Q$ value, say $Q^\ast$, HB2 collides with PB1:  $ Q^\ast = (1- a^2)$. At $Q=Q^\ast$, the AD state, and thus, the AD-OD transition cease to take place. Figure~\ref{fig1}~(b) shows this scenario for $a=0.35$, and $Q=Q^\ast=0.8775$. Now, for $Q>Q^\ast$, ${\epsilon}_{HB2} > {\epsilon}_{PB1}$, i.e., HB2 point moves to the right hand side of PB1; subsequently, the IHSS now becomes stable at $\epsilon_{HBS}$ through a subcritical Hopf bifurcation, where 
\begin{equation}
\label{epsahb3}
{\epsilon}_{HBS} = \frac{1}{\sqrt{(1-Q)}}.
\end{equation}
This is derived from the eigenvalues of the nontrivial fixed point (${x_1}^\ast$, ${y_1}^\ast$, $-{x_1}^\ast$, $-{y_1}^\ast$) given by:
\begin{subequations}
\label{ntlambda}
\begin{align}
\label{ntlambda1-2}
{\lambda}_{1,2} &= \frac{{-b_1}^\ast \pm \sqrt{{{b_1}^\ast}^2-4{c_1}^\ast}}{2},\\
\label{ntlambda3-4}
{\lambda}_{3,4} &= \frac{{-b_2}^\ast \pm \sqrt{{{b_2}^\ast}^2-4{c_2}^\ast}}{2}.
\end{align}
\end{subequations}
Where, ${b_1}^\ast = \epsilon - a(1-{x_{1}^{\ast}}^2)$, ${c_1}^\ast = 1 + 2a{x_{1}^{\ast}}{y_{1}^{\ast}} - a\epsilon(1-{x_{1}^{\ast}}^2)$, ${b_2}^\ast = \epsilon(1-Q) - a(1-{x_{1}^{\ast}}^2)$, ${c_2}^\ast = 1 + 2a{x_{1}^{\ast}}{y_{1}^{\ast}} - a\epsilon(1-Q)(1-{x_{1}^{\ast}}^2)$. Using \eqref{epsahb3}, for $Q=0.95$, we get $\epsilon_{HBS}\approx4.472$ that matches with Fig.\ref{fig1}~(c).

The second nontrivial fixed point $({x_1}^\dagger$, ${y_1}^\dagger$, ${x_1}^\dagger$, ${y_1}^\dagger)$ that was created at $\epsilon_{PB2}$ becomes stable through a subcritical pitchfork bifurcation at $\epsilon_{PBS}$: 
\begin{equation}\label{pbs}
{\epsilon}_{PBS} = \frac{2-Q}{2a(1-Q)^2}.
\end{equation}
This is derived from the eigenvalues corresponding to $({x_1}^\dagger$, ${y_1}^\dagger$, ${x_1}^\dagger$, ${y_1}^\dagger)$, which are same as \eqref{ntlambda} except now the ``$\ast$" signs are replaced by ``$\dagger$" sign. From Fig.\ref{fig1}(a) we have, $\epsilon_{PBS}\approx4.956$, that exactly matches with Eq.\eqref{pbs}. For $\epsilon>\epsilon_{PBS}$,  stable IHSS (OD) (i.e., $x^{\ast}_{1}=-x^{\ast}_{2}$)  coexists with the {\it nontrivial} AD (NT-AD) state (i.e., $x_1^\dagger=x_2^\dagger$) [shaded (cyan) region in Fig.~\ref{fig1} (a)]. The attribute {\it nontrivial} is used because it results from the stabilization of {\it nontrivial} HSSs ($x^\dagger,y^\dagger$). Further, the observed NT-AD is {\it unique} because unlike conventional AD, the NT-AD state is born via a subcritical pitchfork bifurcation. Also, contrary to AD, the NT-AD state is completely destroyed by parameter mismatch \cite{tanod}.
\begin{figure}
\includegraphics[width=.48\textwidth]{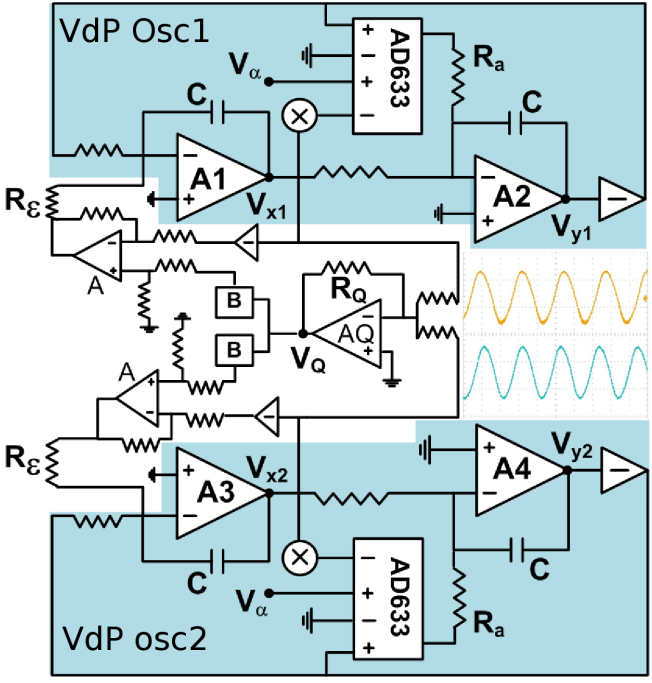}
\caption{\label{ckt} (Color online) Experimental circuit diagram of the mean-field coupled VdP oscillators. A, A1-A4, and AQ are realized with TL074 op-amps. All the unlabeled resistors have value $R=10$~k$\Omega$. C=10 nF, $R_a=286\Omega$, $V_\alpha=0.1$ v. Box denoted by ``B" are op-amp based buffers; inverters are realized with the unity gain inverting op-amps. $\otimes$ sign indicates squarer using AD633. Inset (in the middle part) shows the  oscillation from the uncoupled VdP oscillators: upper trace (yellow) $V_{x1}$, lower trace (cyan) $V_{x2}$ ($y$-axis:10 v/div, $x$-axis:250 $\mu$s/div).}
\end{figure}
The whole bifurcation scenario in the  $\epsilon-Q$ parameter space is  shown in Fig.~\ref{fig1} (d) for $a=0.35$. We can see that, with increasing $Q$, at $Q=0.8775$, collision of HB2 with PB1 destroys the AD-OD transition. It also shows the coexisting region of NT-AD and OD. 
\begin{figure}
\includegraphics[width=\columnwidth]{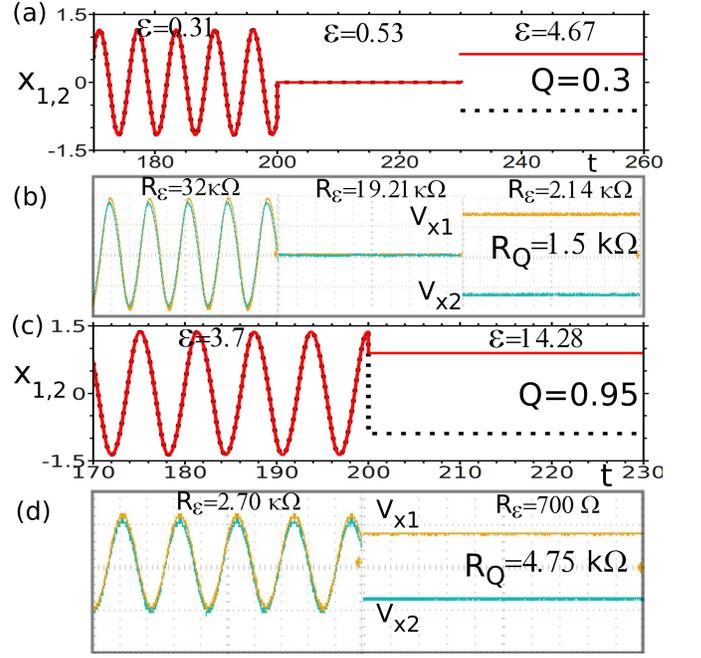}
\caption{\label{expt} (Color  online)  Experimental real time traces [(b) and (d)] of $V_{x1}$ and $V_{x2}$ along with the numerical time series plots [(a) and (c)]  of $x_1$ and $x_2$. (a) $Q=0.3$ (b) $R_Q=1.5$ k$\Omega$: complete synchronized limit cycle at $R_\epsilon=32$ k$\Omega$ ($\epsilon=0.31$), AD at $R_\epsilon=19.2$ k$\Omega$ ($\epsilon=0.53$) , and OD  at $R_\epsilon=2.14$ k$\Omega$ ($\epsilon=4.67$). (c) $Q=0.95$ (d) $R_Q=4.75$ k$\Omega$ : complete synchronized limit cycle at $R_\epsilon=2.70$ k$\Omega$ ($\epsilon=3.70$) and OD state $R_\epsilon=700$ $\Omega$ ($\epsilon=14.28$). [(b), (d): $y$-axis: 10 v/div, $x$-axis: 250 $\mu$s/div].}
\end{figure}

Next, we implement the coupled system in electronic circuit. Figure~\ref{ckt} shows the electronic circuit diagram of two mean-field coupled Van der Pol oscillators. Shaded (blue) regions in the upper and lower portions represent the individual VdP oscillators \cite{vdpckt}. We use TL074 (quad JFET) op-amps, and AD633 analog multiplier chips (having differential inputs); output of the multiplier is scaled by a factor of $0.1$. $\pm15$ v power supplies are used; resistors (capacitors) have $\pm5\%$ ($\pm1\%$) tolerance. The unlabeled resistors have value $R=10$~k$\Omega$. The op-amp AQ is used to generate the mean-field: $V_Q=-\frac{2R_Q}{R}\sum_{j=1}^{2}\frac{V_{xj}}{2}$, which is subtracted from $V_{x1,2}$ using op-amps denoted by A. One can see that $R_\epsilon$ determines the coupling strength, and $R_Q$ determines the mean-field density. The voltage equation of the circuit can be written as:
\begin{subequations}\label{ckteqn}
\begin{align}
CR\dot{V}_{xi}&=V_{yi}+\frac{R}{R_\epsilon}\left[\frac{2R_Q}{R}\sum_{j=1}^{2}\frac{V_{xj}}{2}-V_{xi}\right],\\
CR\dot{V}_{yi}&=\frac{R}{R_a}\left(V_\alpha-\frac{V_{xi}^2}{10}\right)\frac{V_{yi}}{10}-V_{xi}.
\end{align}
\end{subequations}
Here $i=1,2$. Eq.~\eqref{ckteqn} is normalized with respect to $CR$, and thus now becomes equivalent to Eq.~\eqref{system} for the following normalized parameters: $\dot{u}=\frac{du}{d\tau}$, $\tau=t/RC$, $\epsilon=\frac{R}{R_\epsilon}$, $Q=\frac{2R_Q}{R}$, $a=\frac{R}{100R_a}$, $10V_\alpha=1$, $x_i=\frac{V_{xi}}{V_{sat}}$, and $y_i=\frac{V_{yi}}{V_{sat}}$. $V_{sat}$ is the saturation voltage of the op-amp. In the experiment we take $V_\alpha=0.1$ v, and $C=10$ nF; we choose $a=0.35$ by taking $R_a=286$~$\Omega$ [using a precision potentiometer (POT)]. We vary the coupling strength $\epsilon$, and mean-field density $Q$ by varying $R_\epsilon$ and $R_Q$, respectively (using precision POTs). For the uncoupled case, the individual oscillations have a frequency of $1.7$ kHz, and are shown in Fig.~\ref{ckt} (inset). Next, at first we fixed $Q=0.3$ by taking $R_Q=1.5$~k$\Omega$, and vary $R_\epsilon$. With the  increasing coupling strength (i.e., decreasing $R_\epsilon$) we observed the transition from limit cycle (complete synchronized) state to AD at $R_\epsilon=30.9$~k$\Omega$, and then a transition from AD to OD state at $R_\epsilon=3.8$~k$\Omega$. In Fig.~\ref{expt}~(b), using the experimentally obtained snapshots of the waveforms [with a digital storage oscilloscope (Tektronix TDS2002B, 60 MHz, 1 GS/s)], we demonstrate different dynamical behaviors for the following parameter values: complete synchronized limit cycle at $R_\epsilon=32$~k$\Omega$, AD at $R_\epsilon=19.21$~k$\Omega$, and OD  at $R_\epsilon=2.14$ k$\Omega$. For the comparison purpose, we also show the numerical results (using fourth order Runge-Kutta method with step-size 0.01) taking $\epsilon$ values that are equivalent to $R_\epsilon$ (note that $\epsilon=\frac{R}{R_\epsilon}$); Fig.\ref{expt}~(a) shows this with limit cycle (for $\epsilon=0.31$), AD (for $\epsilon=0.53$), and OD (for $\epsilon=4.67$). It can be seen that the numerical and experimental results are in close agreement with each other. As we increase $R_Q$, the AD region reduces; for $R_Q\ge4.32$ k$\Omega$, no AD occurs and the limit cycle state directly transits into OD state beyond a certain coupling strength. This is in agreement to the theory that for $Q>Q^\ast$ (=0.8775) no AD-OD transition takes place. Note the close proximity between $Q^\ast$ and experimental value of $Q^\ast$, i.e., $Q^{\ast}_{expt}=\frac{2R_{Q}^{\ast}}{R}=0.864$. 
\begin{figure}
\includegraphics[width=.4\textwidth]{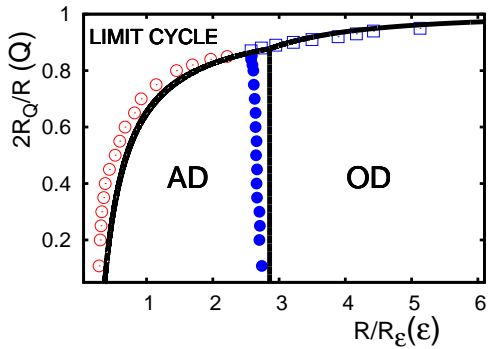}
\caption{\label{fig.4} (Color online) Experimental Phase diagram in the $\frac{R}{R_\epsilon}$ ($\epsilon$)---$\frac{2R_Q}{R}~(Q)$ space. Open circle: experimental transition points from limit cycle to AD; solid circle: experimental transition points from AD to OD; open square: experimental transition points from limit cycle to OD. Theoretical curves (line) are also shown, which closely match with the experimental points.}
\end{figure}
Next, we take $Q=0.95$ ($>Q^\ast$) by taking $R_Q=4.75$~k$\Omega$. Here, in accordance with the theory, we observed direct transition from limit cycle (complete synchronized) to the OD state (at $R_\epsilon=1.95$~k$\Omega$); Fig.~\ref{expt}(d) shows this scenario: limit cycle (at $R_\epsilon=2.7$~k$\Omega$) and OD state (at $R_\epsilon=700$ $\Omega$). Fig.~\ref{expt}(c) shows the same in numerical simulation having limit cycle (at $\epsilon=3.7$), and OD (at $\epsilon=14.28$). We repeat the experiment for a large number of values of $R_Q$ and note the $R_\epsilon$ values where AD, OD, and AD-OD transition occur. To represent the whole experimental scenario, we plot the experimental phase diagram in $\frac{R}{R_\epsilon}$ ($\epsilon$)---$\frac{2R_Q}{R}~(Q)$ space (Fig.\ref{fig.4}). Theoretically obtained curves are also plotted in the same graph. It is noteworthy that the experimental points are in close proximity to the theoretical curves. The slight deviation from the theoretical result occurs due to the inherent parameter fluctuation in electronic circuit, and also the possible parameter mismatches present between the oscillators. We further note that, due to this inherent parameter mismatch, we could not observe the NT-AD state, which is in agreement with the findings of \cite{tanod} that any parameter mismatch destroys the NT-AD state .    

In conclusion, we have experimentally observed the transition from amplitude death to oscillation death state in mean-filed coupled limit cycle oscillators. We have chosen the paradigmatic Van der Pol oscillators coupled via mean-field diffusion, and implemented the system in electronic circuit. By changing the coupling strength for a fixed mean-field parameter, we have experimentally observed the transition from AD to OD if the  mean-field parameter has a value less than a threshold value. Beyond that threshold value, no AD occurs, and limit cycle oscillation directly transforms into a OD state. We have corroborated the experimental results by suitable theory and bifurcation analysis. We believe that apart from electronic circuits the AD-OD transition scenario can be observed experimentally in laser and neuronal systems that may reveal the practical application of this transition in those systems. 
\providecommand{\noopsort}[1]{}\providecommand{\singleletter}[1]{#1}%
\end{document}